\documentclass[12pt]{article}
\usepackage[pdftex]{color}
\usepackage{amsmath,amsthm,amssymb}
\usepackage{wasysym}
\usepackage{graphicx}
\usepackage{color}%
\usepackage[onehalfspacing]{setspace}
\usepackage{lipsum}
\usepackage[lmargin=2.5cm,rmargin=2.5cm,tmargin=3cm,bmargin=3cm]{geometry}
\usepackage{multirow}
\usepackage{hyperref}
\hypersetup{hidelinks}
\usepackage{cleveref}
\usepackage{accents}
\usepackage[ruled,vlined,linesnumbered]{algorithm2e}
\usepackage[authoryear,sort&compress]{natbib}
\usepackage{scalerel}
\usepackage{stackengine}
\usepackage{booktabs}
\usepackage[ruled,vlined,linesnumbered]{algorithm2e}
\stackMath
\definecolor{color}{rgb}{0.8500, 0.3250, 0.0980} 

\begin{document}
\begin{center}
{\rm \bf \Large{Stochastic process model of rough surface contact}}
\end{center}

\begin{center}
{\bf Yang Xu} \\
School of Mechanical Engineering, Hefei University of Technology, Hefei, 230009, China \\
yang.xu@hfut.edu.cn\\
+86 173~4406~1443
\end{center}

\begin{center}
{\bf Yunong Zhou} \\
Department of Civil Engineering, Yangzhou University, Yangzhou, 225127, China \\
yunong.zhou@yzu.edu.cn\\
+86 138~5272~2693
\end{center}

\begin{abstract}
The stochastic process model of rough surface contact, widely known as Persson's theory of contact, serves as a representative multi-scale model that has been extensively applied across various fields of tribology. In this chapter, we briefly introduce the background of the development of Persson's theory of contact. We thoroughly discuss Persson's theory for purely normal elastic contact, with a special focus on solving the probability density of the contact pressure and the interfacial gap using partial differential equations. Subsequent applications of these fundamental results in addressing more complex interfacial properties in other fields of tribology are also examined. Finally, several recommendations regarding future studies of Persson's theory are proposed. This review article is expected to assist researchers in quickly familiarizing themselves with the current state of the art of Persson's theory and to attract more attention from tribologists and solid mechanicians, thereby contributing to the development and application of Persson's theory of contact.
\end{abstract}

{\bf Keywords:} 
Persson's theory of contact; rough surface contact; stochastic process; multi-scale model; adhesion; friction; sealing; wear; inelastic contact; elastic contact

{\bf Keypoints:}
\begin{itemize}
\item Background introduction of multi-scale models including Persson's theory of contact
\item Review of fundamentals of Persson's theory in purely normal elastic contact problems
\item A brief introduction of Persson's theory of contact in various various fields of tribology
\item Recommendations for future studies of Persson's theory
\end{itemize}

\section{Introduction}

Natural and machined surfaces exhibit some degree of uneven surface texture, resulting in various tribological phenomena, such as earthquakes \citep{van2018subduction}, Joule heating at electrical contacts \citep{sui2023modeling}, and squeaking in ceramic-on-ceramic bearings used in total hip arthroplasty \citep{brockett2013squeaking}. By observing the same surface texture at different length scales, one can identify a sequence of similarly repeating random rough surface topographies, characterized by an increasing surface height-to-wavelength ratio \citep{Gujrati18}. This phenomenon is known as the multi-scale nature of surface texture and is often referred to as fractals \citep{Mandelbrot84}. Since the seminal work of \citet{archard1957elastic}, numerous efforts have been dedicated to integrating the multi-scale nature of rough surfaces with contact mechanics theory to explore the influence of geometric scale on interactions between rough surfaces. These efforts include investigations into the real area of contact, interfacial contact stiffness, mean interfacial gap, and other interfacial mechanical parameters. In this section, the randomness and stochastic nature of rough surface topography, which are essential to Persson’s theory of contact, will first be introduced. Then, a brief overview of the historical development of multi-scale models will be provided to emphasize the background of the creation and evolution of Persson’s theory of contact.

\subsection{Randomness and stochastic nature of rough surface topography}

Unlike conventional contact mechanics problems, where surfaces in contact are smooth (for instance, the parabolic surface in Hertzian contact theory \citep{johnson1987contact}), the topographies of surfaces in reality are rough and are often characterized as a random or stochastic process. The terms ``random" and ``stochastic" are frequently used interchangeably, even in the field of statistics; however, their differences are clearly outlined below to distinguish multi-scale contact models from single-scale contact models. 

Since the contact of rough surfaces is a localized interfacial phenomenon with a negligible ratio of interfacial gap to lateral dimension within the nominal contact area, the rough surface topography under interaction, denoted by $h(x, y)$, is typically regarded as nominally flat. This topography can either be a measured point cloud or its digital twin. Following Nayak's random process theory \citep{Nayak71}, the statistics of a random rough surface are commonly characterized by a fixed joint probability density function (PDF) in terms of rough surface height and its derivatives with respect to in-plane coordinates: 
\begin{equation}\label{eq:PDF_topograph}
P\left(h, \frac{\partial h}{\partial x}, \frac{\partial h}{\partial y}, \frac{\partial^2 h}{\partial x^2}, \frac{\partial^2 h}{\partial x \partial y}, \frac{\partial^2 h}{\partial y^2}, \cdots \right).
\end{equation}
The PDF-based characterization of rough surface topographies is extensively utilized in multi-asperity contact models \citep{Greenwood66, Bush75, carbone2008asperity, xu2014statistical}. The measured rough topography is bandwidth-limited, as the upper and lower cut-off wavelengths ($\lambda_{\text{l}}$ and $\lambda_{\text{s}}$) are strongly dependent on the type of measurement technique. The upper cut-off wavelength ($\lambda_{\text{l}}$) is directly related to the length of the sampling area, which is constrained by the maximum travel distance of the stylus or the field of view of the optical sensor. The lower cut-off wavelength ($\lambda_{\text{s}}$) is determined by the sampling resolution, which can vary from several hundred microns to several angstroms \citep{Gujrati18}. The rough surface contact model that utilizes Eq. \eqref{eq:PDF_topograph} for characterizing the random rough topography is often referred to as a single-scale model due to its fixed magnification $\zeta=\lambda_{\text{l}}/\lambda_{\text{s}}$. 

\begin{figure}[h!]
  \centering
  \includegraphics[width=12cm]{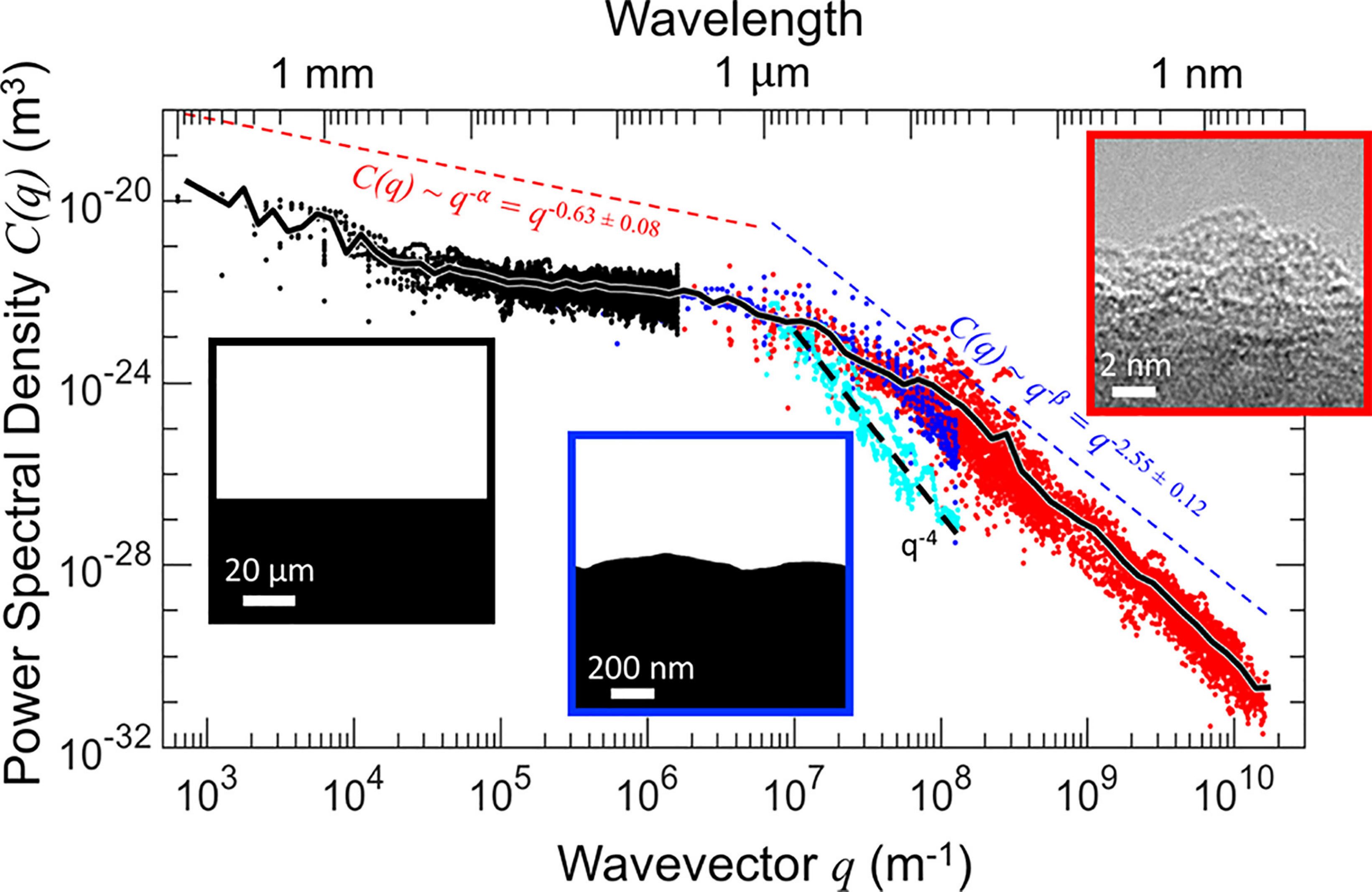}
  \caption{Power spectral density of a rough topography across eight orders of magnitude in length scale. Reproduced with permission from \citet{Gujrati18}, Copyright 2018 American Chemical Society.}\label{fig:Fig_2}
\end{figure}

A real surface texture with finite dimensions has a fixed $\lambda_{\text{l}}$ and an infinitesimally small $\lambda_{\text{s}}$. Rough surface topographies with a decreasing lower cut-off wavelength (or, equivalently, an increasing magnification $\zeta$) can be classified as stochastic processes. Unlike a random process with a fixed PDF, the probability distribution of the stochastic process evolves with increasing $\zeta$. Although the geometry of rough surface topography is mathematically indeterministic, certain groups of rough topographies can be characterized deterministically in terms of their power spectral density (PSD). A recent Surface-Topography Challenge \citep{pradhan2025surface}, launched by Tevis Jacobs, has provided convincing evidence that the radially averaged PSD-to-wavenumber relations measured over similar surface samples by various measurement techniques converge into a single curve. Similar phenomena have been reported by \citet{Sayles78} and \citet{Gujrati18} (see Fig. \ref{fig:Fig_2}). These observations strongly indicate that the multi-scale nature of the rough topography can be deterministically characterized using a bandwidth-limited PSD, or equivalently, the spectrum. This characterization has been utilized to represent rough surface topography in several multi-scale models, such as the Jackson and Streator model \citep{jackson2006multi} and Persson’s theory of contact \citep{Persson01}. 

\subsection{Historical development of multi-scale contact models}
\begin{figure}[h!]
  \centering
  \includegraphics[width=16cm]{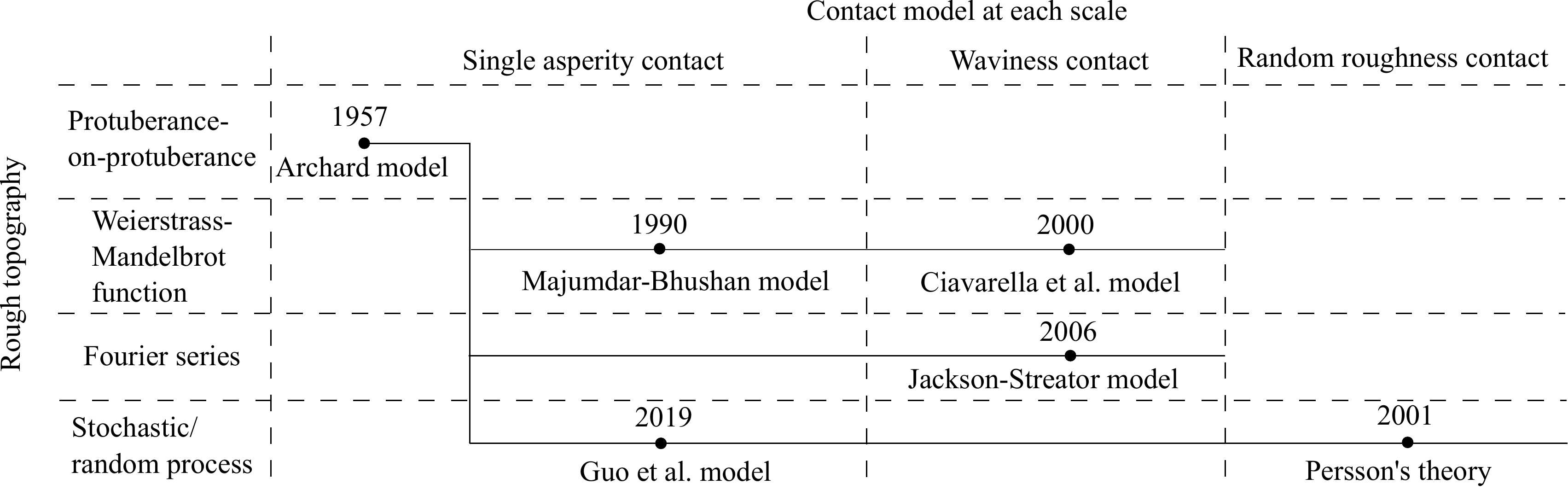}
  \caption{Graphical illustration of the historical development of the multi-scale models of rough surface contact}\label{fig:Fig_3}
\end{figure}
A graphical illustration of the historical development of multi-scale models, highlighting some well-known examples, is shown in Fig. \ref{fig:Fig_3}. It is commonly believed that \citet{archard1957elastic} developed the first multi-scale model, long before the invention of modern profilometry for multi-scale topographical analysis and the concept of fractals. Archard’s model characterizes the purely normal interaction between a rigid flat and a rough sphere featuring a “protuberances-on-protuberances” structure, where hemispherical surfaces of smaller radius lie uniformly on top of the larger hemispherical surfaces. The interaction between the outermost asperities and the rigid flat can be characterized by Hertzian circular contact theory. The concept of asperity contact adopted in the Archard model has a significant influence on the subsequent development of multi-scale models. This broccoli-shaped structure exhibits self-similarity as $\zeta$ increases; however, it cannot accurately represent the true rough topography. 

The next cornerstone, commonly referred to as the Majumdar-Bhushan (MB) model \citep{majumdar1990}, combines the single asperity contact model and the fractal characterization of rough topography, following its introduction by Mandelbrot to characterize 3D topographies in the 1970s \cite{Mandelbrot84}. Multi-scale topography is achieved through the superposition of sinusoidal components, as described by the Weierstrass-Mandelbrot function \citep{berry1980weierstrass,yan1998contact}. The bearing model is utilized to truncate the rough topography, allowing for the identification of contacting asperities and enabling the determination of the real contact area. The multi-asperity contact (single-scale) model has been independently extended by \citet{guo2019magnification} and \citet{jackson2023asperity} to account for the multi-scale nature of the surfaces. It is assumed that the surface topography possesses a deterministic PSD, allowing the corresponding moments to be determined in terms of the magnification. The multi-scale Greenwood-Williamson (GW) model can predict the scale-dependent real area of contact, interfacial gap, and other related parameters. 

In 2000, \citet{ciavarella2000linear} developed a multi-scale contact model under the plane strain condition based on Archard's concept of ``protuberances-on-protuberances." This model inherits the fractal characterization of the rough profile through the Weierstrass-Mandelbrot function, while the contact is modeled using the Westergaard solution \citep{johnson1987contact}, which effectively incorporates the influence of neighboring asperity contacts and enables comprehensive solutions ranging from initial contact to complete contact. Moreover, \citet{ciavarella2000linear} formulated the evolution of the PDF of the contact pressure due to an incremental change of the magnification using an integral, allowing the model of Ciavarella et al. to effectively simulate the response of the contact pressure distribution as additional rough surface components with smaller wavelengths are progressively included. We will demonstrate later that this approach is essentially the same as that used in Persson’s theory of contact \citep{Persson01}. 

One year after the model of Ciavarella et al., \citet{Persson01} published his seminal work on tire-road interaction, in which he ingeniously developed a differential equation in Appendix B to characterize the evolution of the probability density function (PDF) of the contact pressure with magnification. Unlike all previously published works, the rough surface topography in Persson's theory is uniquely characterized by the PSD. In 2007, \citet{jackson2006multi} adapted Archard’s model by replacing the asperity contact and broccoli-shaped surface with the waviness contact model and the rough topography characterized by a Fourier series. The sinusoidal waviness at each scale can be effectively decomposed using the fast Fourier transform applied to the measured topographical data. For a given normal load and magnification, the real area of contact and interfacial gap can be determined iteratively by progressively stacking the sinusoidal waviness with smaller wavelengths.

Persson's theory of contact is commonly referred to as a unique interfacial modeling technique that deterministically characterizes the evolution of the PDF of stochastic variables, including contact pressure \citep{Persson01,Persson01b} and interfacial gap \citep{Almqvist11,xu2024stochastic}. These results serve as the foundation for calculating the global interfacial properties, such as the mean interfacial gap \citep{Persson07}, real area of contact \citep{Manners06}, elastic strain energy \citep{Persson08}, wear \citep{persson2025rubber}, friction \citep{Persson01}, etc. It has been widely applied in almost every aspect of tribological studies and beyond, including applications in seal \citep{persson2016leakage}, lubrication \citep{scaraggi2018influence}, tire-road interaction \citep{Persson01}, and rock mechanics \citep{lang2015hydraulic,kling2018numerical}. 

The review of Persson's theory will be organized as follows: Section 2 discusses the rough surface topography used in Persson's theory; Section 3 provides a brief introduction to Persson's theory of contact concerning purely normal, elastic, non-adhesive contact; Sections 4 and 5 present a discussion of Persson's theory related to non-adhesive, inelastic contact and adhesive contact, respectively; Section 6 reviews applications of Persson's theory of contact in other fields of study, followed by a discussion of suggested future works in Section 7. 

\section{Rough surface topography}
The topographies of certain groups of rough surfaces can be characterized deterministically by a piecewise PSD as follows \citep{Jacobs17, persson2004nature}: 
\begin{equation}\label{eq:PSD_piecewise}
C(q) = 
\begin{cases}
C_0 q_{\text{r}}^{-2(1 + H)} ~~~ q \in [q_{\text{l}}, \min(q_{\text{r}}, \zeta q_{\text{l}})), \\
C_0 q^{-2(1 + H)} ~~~ q \in [\min(q_{\text{r}}, \zeta q_{\text{l}}), \zeta q_{\text{l}}], \\
0 ~~~ \text{elsewhere},
\end{cases}
\end{equation}
where $q_{\text{l}} = 2 \pi/\lambda_{\text{l}}$, $q_{\text{r}}$, and $\zeta q_{\text{l}} = 2 \pi /\lambda_{\text{s}}$ are the lower cut-off, roll-off, and upper cut-off wavenumbers, respectively; $\lambda_{\text{l}}$ and $\lambda_{\text{s}}$ are the longest and shortest wavelengths, respectively. $H \in [0, 1]$ is the Hurst exponent; $C_0$ is a constant. It has been confirmed that Eq. \eqref{eq:PSD_piecewise} can characterize PSDs of many natural and machined surface textures \citep{Sayles78,Gujrati18,persson2004nature}. It is commonly assumed, though not often explicitly stated, that the rough surface topography used in Persson's theory of contact is \emph{random}, \emph{bandwidth-limited}, \emph{self-affine}, and \emph{isotropic}.  The randomness of the rough topography arises from the fact that the phase angle of each sinusoidal component is random \citep{wu2000simulation}. The bandwidth-limited nature of the surface arises because the PSD in Eq. \eqref{eq:PSD_piecewise} is truncated at the lower and upper cut-off wavenumbers. Self-affinity of the rough surface is ensured when $H < 1$ to avoid the unrealistic self-similar feature in which all three dimensions of the rough surface topography scale uniformly. Isotropy is commonly assumed in the classic Persson's theory of contact, where the two-dimensional PSD must be strictly axisymmetric, i.e., $C(q_x, q_y) = C(q)$, $q = \sqrt{q_x^2 + q_y^2}$. This implies that the statistical properties of the rough topography are uniform in all in-plane directions \citep{Nayak71}. Anisotropic rough topography can also be employed in Persson's theory without restriction \citep{carbone2009contact, ciavarella2024strongly}.

According to the law of large numbers, the rough surface height follows a Gaussian distribution for $\zeta > 1$, where the PDF is 
\begin{equation}
P(h, \zeta) = \frac{1}{\sqrt{2 \pi \sigma^2(\zeta)}} \exp \left( -\frac{h^2}{2 \sigma^2(\zeta)} \right),
\end{equation}
where $\sigma^2(\zeta) = 2 \pi \int_{q_{\text{l}}}^{\zeta q_{\text{l}}} q C(q) dq$ represents the magnification-dependent variance of surface height. This condition is often overlooked in the validation of Persson's theory against various numerical models and experimental results, where the surface height is expected to follow a Gaussian distribution \citep{Xu17}. As $\zeta$ increases by an incremental step $\Delta \zeta$, the evolution of $P(h, \zeta)$ can be described by the following integral:
\begin{equation}
P(h, \zeta + \Delta \zeta) = \int_{-\infty}^{\infty} P(h, \zeta + \Delta \zeta |h', \zeta) P(h', \zeta) dh'.
\end{equation}
The transition PDF is 
\begin{equation}
P(h, \zeta + \Delta \zeta|h', \zeta) = \frac{1}{\sqrt{2 \pi} \Delta \sigma} \exp \left( -\frac{(h - h')^2}{2 (\Delta \sigma)^2} \right),
\end{equation} 
where $\Delta \sigma = \sigma(\zeta + \Delta \zeta) - \sigma(\zeta)$.

\section{Purely normal elastic non-adhesive contact}
Persson's theory of contact was initially applied to solve the purely normal contact problem. It aims to determine the evolution of the PDF of contact pressure with magnification. Other interfacial properties, such as the relative contact area, elastic strain energy, average interfacial gap, and the PDF of the interfacial gap, can be deduced either directly or indirectly from this fundamental solution. This theory has been extensively developed within this fundamental category, and the fruitful results have established a basis for its application in more complex fields. In this section, the core concepts of Persson's theory in elastic, non-adhesive contact will be revisited in detail. 

\subsection{Problem statement}
\begin{figure}[h!]
  \centering
  \includegraphics[width=16cm]{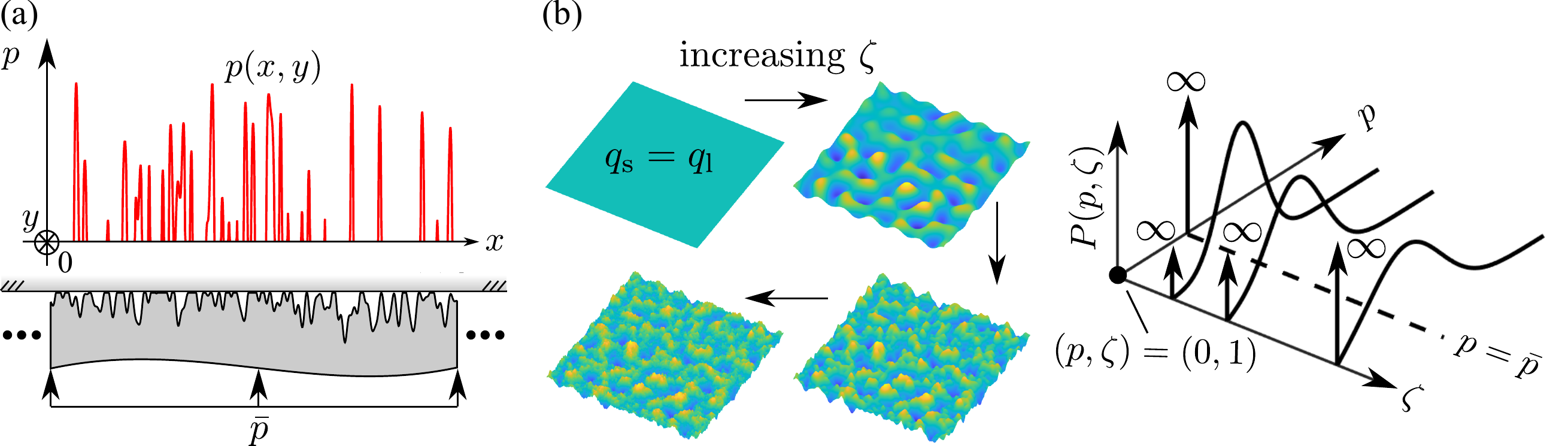}
  \caption{Schematic of (a) purely normal contact problem and (b) evolution of contact pressure with magnification. Reproduced with permission from \citet{xu2024persson}, Copyright 2024 Elesvier.}\label{fig:Fig_4}
\end{figure}

The purely normal contact can be analyzed using Persson's theory of contact, which is stated as follows: Contact occurs between two uncorrelated rough surface topographies, namely $h_1(x, y)$ and $h_2(x, y)$, $(x, y) \in \mathbb{R}^2$. According to \citet{barber2003bounds}, the contact between two rough surfaces is equivalent to an indentation problem, in which a composite, rigid surface ($h(x, y) = h_1(x, y) + h_2(x, y)$) is in contact with an elastic half-space. This scenario is analogous to a flattening problem, in which an elastic, composite, rough surface is flattened by a rigid flat (see Fig. \ref{fig:Fig_4}). Unless otherwise stated, the contact problems discussed below are indentation problems, in which the elasticity of the rough surface is characterized by the plane strain modulus, $E^* = 1/[(1 - \nu_1^2)/E_1 + (1 - \nu_2^2)/E_2]$, where $E_i$ and $\nu_i$, $i = 1, 2$, are Young's modulus and Poisson's ratio of the two contacting bodies in the original problems, respectively. The half-space is spatially fixed, while the rigid flat is subjected to a uniform normal traction at the far end. The contact pressure and interfacial gap distribution over the interface are $p(x, y)$ and $g(x, y)$, respectively. 

\subsection{Evolution of PDF of contact pressure with magnification}
\citet{Persson01} derived a diffusion equation to describe the evolution of the PDF of the contact pressure with magnification in Appendix B, utilizing a general viscoelastic constitutive law. The derivation is overly simplified with only five steps. Following this work, a simplified variant has been developed, focusing on elastic contact \citep{persson2002elastic}. The major components of Persson's theory, as presented by \citet{persson2002elastic} and other references, are as follows: Let $P(p, \zeta)$ be the magnification-dependent PDF of the random contact pressure $p$ over the entire contact interface. The relative contact area, defined as the ratio of the real area of contact to the nominal contact area, is denoted by $A^*$. The contact pressure PDF can be decomposed into \citep{Persson18,Xu22,xu2024persson}
\begin{equation}\label{eq:PDf_p_general_form}
P(p, \zeta) = (1 - A^*) \delta(p) + P_0(p, \zeta),
\end{equation} 
where $P_0(p, \zeta)$ is the PDF of the contact pressure within the contact area, satisfying the following diffusion equation \citep{Persson01,xu2024persson,Xu2025Revisiting}:
\begin{equation}\label{eq:Diffusion}
\frac{\partial}{\partial \zeta} P_0(p, \zeta) = \frac{1}{2} B_2(\zeta) \frac{\partial^2}{\partial p^2} P_0(p, \zeta). 
\end{equation}
The diffusion coefficient, $B_2(\zeta) = d\text{Var}(p)/d\zeta$, represents the rate of change of the variance of contact pressure, $\text{Var}(p)$, with respect to the magnification \citep{Manners06}. The diffusion coefficient at partial contact ($A^* < 1$) does not have an analytical solution. Assuming $B_2(\zeta)$ at partial contact is the same as that at complete contact, \citet{Persson01} provides the first approximation as
\begin{equation}\label{eq:Diffusion_approximate}
B_2(\zeta) = d\text{Var}(p)/d\zeta \approx dV/d\zeta,
\end{equation}
where $V(\zeta) = \displaystyle{\frac{\pi}{2} (E^*)^2 \int_{q_{\text{l}}}^{\zeta q_{\text{l}}} q^3 C(q) dq}$ represents the variance of contact pressure at complete contact (i.e., $A^* = 1$) \citep{xu2024persson}. A general form of $V(\zeta)$ has been established by Persson for layered contact with finite thickness \citep{persson2012contact}.  

\citet{Wang17} provided an alternative form that describes the evolution of the pressure PDF in response to an incremental change of the magnification:
\begin{equation}\label{eq:CK_equation}
P_0(p, \zeta + \Delta \zeta) = \int_0^{\infty} P_0(p, \zeta + \Delta \zeta|p', \zeta) P_0(p', \zeta) dp'.
\end{equation}
Sequentially applying Eq. \eqref{eq:CK_equation} with a constant or varying $\Delta \zeta$ enables the determination of the PDF of contact pressure at an arbitrary magnification. Equation \eqref{eq:CK_equation} is known as the Chapman-Kolmogorov equation. Its differential form is exactly the same as Eq. \eqref{eq:Diffusion} under specific conditions \citep{xu2024persson}. 

Two fundamental assumptions have been implicitly adopted in Persson's theory of contact: (1) the variation of the random contact pressure with respect to magnification is modeled as a Markov process, and (2) re-entry is not allowed. The first assumption arises from the fact that the evolution of the PDF, as indicated in Eq. \eqref{eq:CK_equation}, depends solely on its most ``recent" magnification history. 
This ``memoryless" feature is a characteristic of a Markov process \citep{xu2024persson}. The second assumption was identified by \citet{Dapp14}. 
It states that once an arbitrary location is out of contact with the magnification $\zeta$, it will remain out of contact with any arbitrarily larger magnification, i.e., 
\begin{equation}
P_0(p, \zeta + \Delta \zeta | p'=0, \zeta) = \delta(p), 
\end{equation}
where $\Delta \zeta > 0$. 

Eq. \eqref{eq:Diffusion} is subjected to the following boundary conditions:
\begin{equation}\label{eq:BC}
P_0(p = 0, \zeta) = 0, ~~~ P_0(p \to \infty, \zeta) = 0.
\end{equation}
The former is known as the absorbing boundary condition, and a rigorous proof has been given by \citet{Persson06} based on the fact that $\frac{\partial}{\partial \zeta} \int_0^{\infty} p P_0(p, \zeta) dp = 0$. The latter is a natural boundary condition that must be satisfied for a well-defined $P_0(p, \zeta)$. The initial condition $P_0(p, \zeta = 1)$ represents the PDF of contact pressure when the surface texture is free of rough texture. A limited number of cases have closed-form solutions for $P_0(p, \zeta = 1)$, including flat-on-flat contact \citep{Persson01,Manners06}, Hertzian circular contact, and cylindrical contact \citep{persson2002elastic,Manners06}. The most commonly used formulation is $P_0(p, \zeta = 1) = \delta(p - \bar{p})$ for flat-on-flat contact. The remaining contact problems depend on numerical methods (e.g., finite element method) to obtain the discretized form of the initial condition. The initial condition, along with $P_0(p = 0, \zeta) = 0$, ensures load equilibrium. 

\begin{figure}[h!]
  \centering
  \includegraphics[width=15cm]{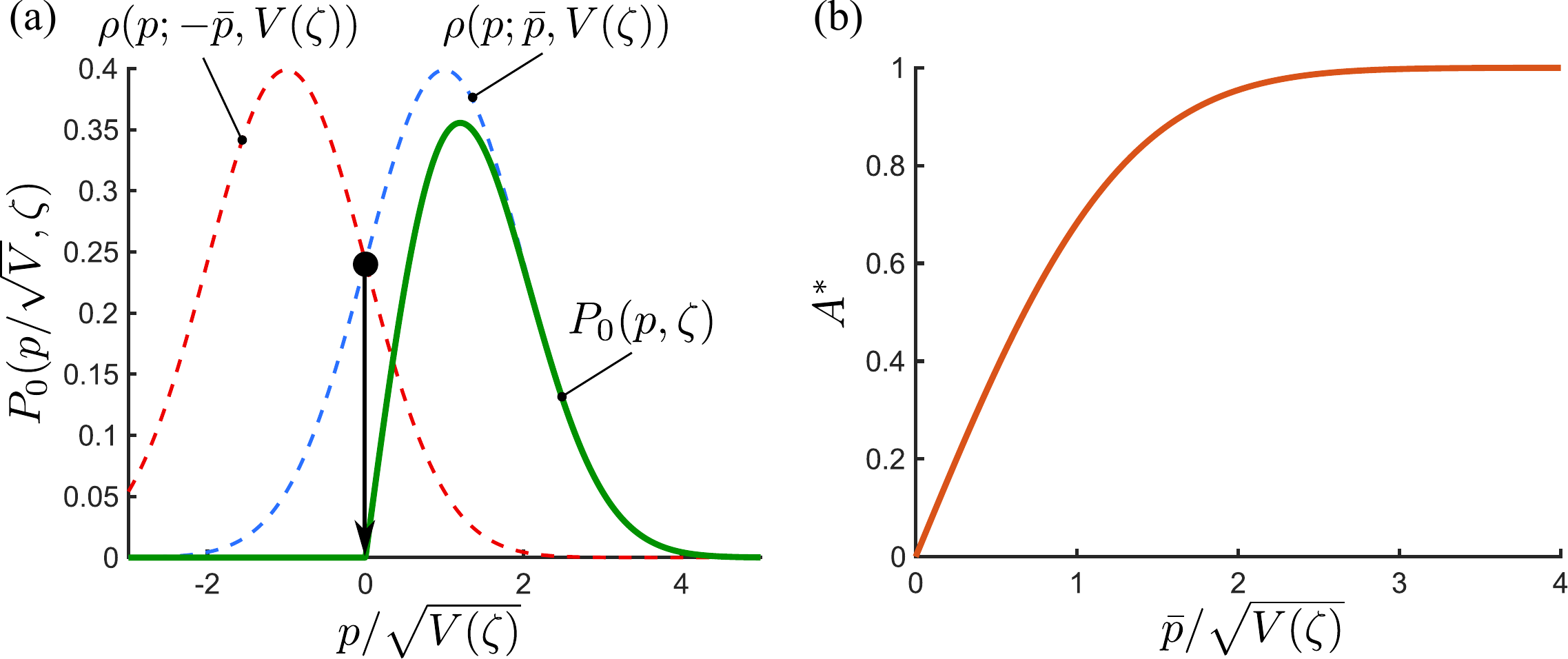}
  \caption{Schematic of (a) $P_0(p, \zeta)$ PDF of contact pressure and (b) $A^*(\bar{p})$ relation; Contact occurs between a nominally flat rigid rough surface and an elastic half-space. The Gaussian distribution $\rho(p; x, V) = \frac{1}{\sqrt{2 \pi V }} \exp \left( -\frac{(p - x)^2}{2 V} \right)$. Reproduced with permission from \citet{xu2024persson}. Copyright 2024 Elsevier. }\label{fig:Fig_5}
\end{figure}

For nominally flat rough surface contact, \citet{Persson01} provided the analytical form of the PDF of the contact pressure as an infinite sum of sines. A simplified closed-form solution has been independently developed by various research groups, including \citet{Yang06}, \citet{Manners06}, and \citet{greenwood2006simplified}. Inspired by the work of Carslaw and Jaeger on heat conduction in solids \citep{Manners06}, the closed-form solution of Eq. \eqref{eq:Diffusion}, subjected to the boundary conditions (Eq. \eqref{eq:BC}) and the initial condition $P_0(p, \zeta = 1) = \delta(p)$, is composed of a Gaussian distribution, from which its mirror image about $p = 0$ is subtracted (see Fig. \ref{fig:Fig_5}(a)):
\begin{equation}\label{eq:pPDF_nominallyflat}
P_0(p, \zeta) = \frac{1}{\sqrt{2 \pi V(\zeta)}} \left( \exp\left[ -\frac{(p - \bar{p})^2}{2 V(\zeta)} \right] - \exp\left[ -\frac{(p + \bar{p})^2}{2 V(\zeta)} \right] \right).
\end{equation}
The diffusion coefficient takes the approximate form provided in Eq. (8). The relative contact area, $A^* = \int_0^{\infty} P_0(p, \zeta) dp$, for nominally flat rough surface contact is expressed in a well-known closed form as follows \citep{persson2004contact, Manners06,Yang06,Persson06}:
\begin{equation}\label{eq:rel_area}
A^* = \text{erf}\left(\bar{p}/\sqrt{2 V(\zeta)}\right),
\end{equation}
where $\text{erf}(x) = \frac{2}{\pi} \int_0^x e^{-t^2} dt$ is the error function (see Fig. \ref{fig:Fig_5}(b)). 

When two interacting surfaces are no longer nominally flat, the contact pressure without roughness, i.e., $P_0(p, \zeta = 1)$, must be determined either analytically or numerically. A general form of $P_0(p, \zeta)$ was initially provided by \citet{persson2002elastic} while a simpler form was given by \citet{Manners06} as follows:
\begin{equation}\label{eq:PDF_p_general}
P_0(p, \zeta) = \frac{1}{\sqrt{2 \pi V(\zeta)}} \int_0^{\infty} P_0(p', \zeta = 1) \left( \exp\left[ -\frac{(p - p')^2}{2 V(\zeta)} \right] - \exp\left[ -\frac{(p + p')^2}{2 V(\zeta)} \right] \right) dp'. 
\end{equation} 
When a surface is nominally flat ($P_0(p, \zeta = 1) = \delta(p - \bar{p})$), Eq. \eqref{eq:PDF_p_general} degrades to a double Gaussian form, as given in Eq. \eqref{eq:pPDF_nominallyflat}. For rough spherical contacts, \citet{persson2002elastic} provided a closed-form solution for $P_0(p, \zeta)$ as an infinite sum of sines, while a simpler form was derived by \citet{Manners06}. For rough cylindrical contacts, the closed form of $P_0(p, \zeta)$ is presented as an infinite sum of sines \citep{persson2002elastic}. 

The application of Persson's theory requires careful selection of $\zeta$. As $\zeta$ approaches infinity, a paradoxical outcome emerges: the relative contact area vanishes asymptotically. This phenomenon, also observed in the multi-scale model of \citet{ciavarella2000linear}, stems from oversimplifications in both the interaction law and the constitutive law. Subsequent studies have resolved this issue by adopting inter-molecular interaction \citep{persson2002elastic, Joe17} and the elastoplastic constitutive law \citep{Persson01b}.

Complementary to Persson's development, \citet{Mueser2008PRL} formulated a rigorous field-theoretical approach for rough contact mechanics. In this framework, the pressure PDF is obtained through a statistical cumulant expansion, which reduces to Persson's theory at the leading order while allowing for the incorporation of higher-order corrections. This formalism not only captures the elastic coupling between asperities but also provides a consistent method to include non-Gaussian tails in the pressure PDF when higher-order moments are considered \citep{Mueser2008PRL,campana2008elastic}.

\subsection{Similarity with multi-asperity contact model}
At light load ranges where $\bar{p}$ and $A^*$ are infinitesimally small, the asymptotic form of $P_0(p, \zeta)$ is \citep{Manners06}
\begin{equation}
P_0(p, \zeta) \to \frac{1}{\sqrt{2 \pi V(\zeta)}} \left( \frac{2 p \bar{p}}{V(\zeta)} \right) \exp\left(-\frac{p^2}{2 V(\zeta)}\right). 
\end{equation}
The asymptotic $A^*(\bar{p})$ is a linear relation \citep{Persson01, Manners06, xu2024persson, ciavarella2006re}:
\begin{equation}\label{eq:A_lowload}
A^* \to  \displaystyle{\sqrt{\frac{2}{\pi}}\ \frac{\bar{p}}{\sqrt{V(\zeta)}} = \kappa \frac{\bar{p}}{E^* \sqrt{\langle |\nabla h |^2\rangle}}}, 
\end{equation}
where $\kappa = \sqrt{8/\pi}$, $\langle |\nabla h|^2 \rangle = \langle (\partial h/\partial x)^2 + (\partial h/\partial y)^2 \rangle$ is the root mean square surface gradient \citep{Yastrebov15}, and $V(\zeta) = \frac{1}{4} (E^*)^2 \langle |\nabla h|^2 \rangle$. The multi-asperity contact models are well-suited for light-load conditions, where the external normal load is balanced by asperity contacts. The Bush, Gibson, and Thomas (BGT) model \citep{Bush75}, which is one of the representative multi-asperity contact models, predicts almost the same asymptotic behavior as that described in Eq. \eqref{eq:A_lowload} except for a different proportionality: $\kappa = \sqrt{2 \pi}$ \citep{hyun2004finite}. 

\citet{xu2014statistical} extended the multi-asperity contact models to heavy load ranges by replacing the Hertzian asperity contact model with a penny-shaped crack model. It has been confirmed that the $A^*(\bar{p})$ relations predicted by the model of \citet{xu2014statistical} and Persson's theory are quantitatively similar \citep{xu2014statistical, greenwood2015almost, Xu17}. \citet{ciavarella2016rough} derived an asymptotic form of $A^*$ based on the model of \citet{xu2014statistical}, which is almost identical to the error functional form of Persson's theory, except for a different proportionality:
\begin{alignat}{3}
\text{\citet{Persson01}} &: ~~~ A^* = 1~- &&\text{erfc}\left(\bar{p}/\sqrt{2 V(\zeta)}\right), \label{eq:Persson_solu} \\
\text{\citet{ciavarella2016rough}} &: ~~~ A^* \to 1 - \frac{3}{4} &&\text{erfc}\left(\bar{p}/\sqrt{2 V(\zeta)}\right), \label{eq:Ciavarella_solu}
\end{alignat}
where $\text{erfc}()$ is the complementary error function. 

\subsection{Fudge factors}

\begin{figure}[h!]
  \centering
  \includegraphics[width=16cm]{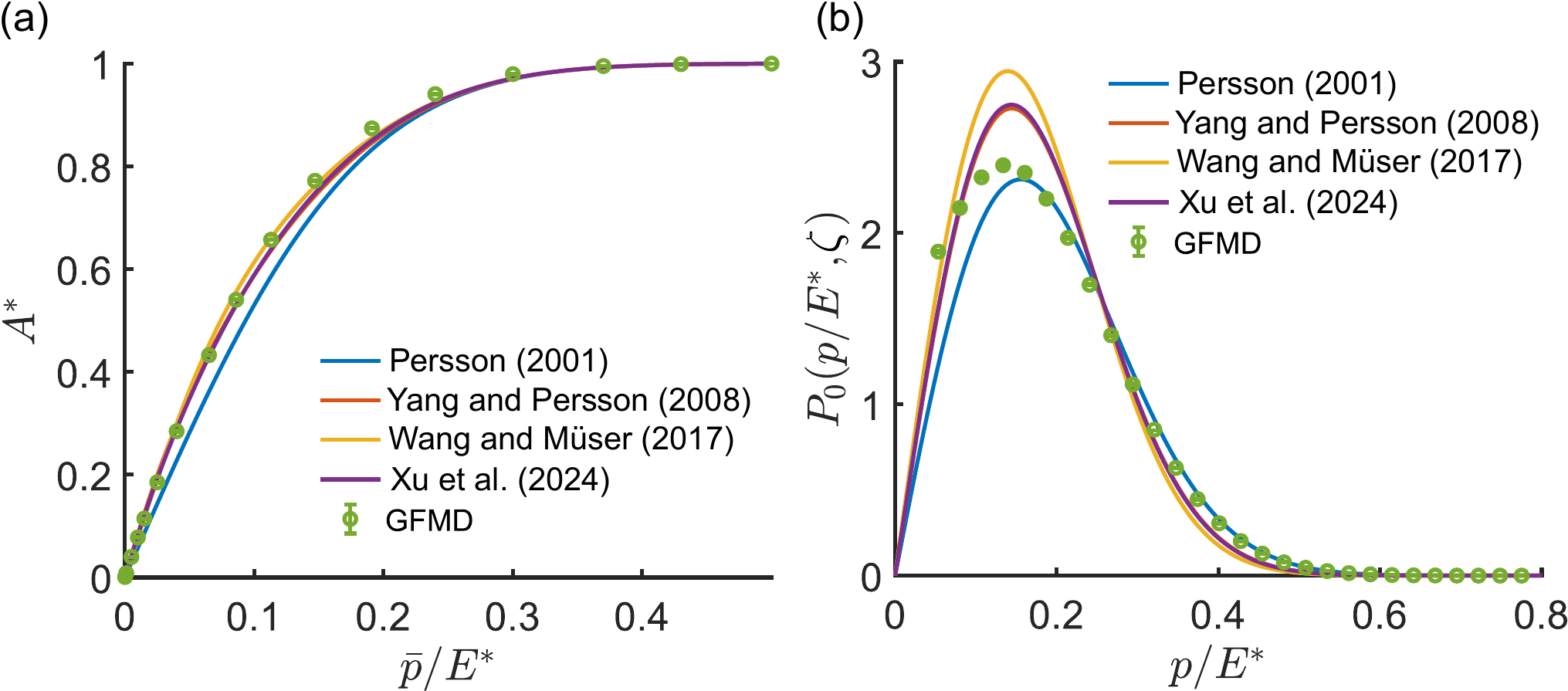}
  \caption{Comparison between $A^*(\bar{p})$ relations predicted by Persson's theory with various fudge factors and Green's Function Molecular Dynamics. Reproduced with permission from \citet{Xu22}. Copyright 2024 Springer Nature.}\label{fig:Fig_6}
\end{figure}

The approximated form in Eq. \eqref{eq:Diffusion_approximate} overestimates the diffusion coefficient. Several numerical studies have shown that the original Persson's theory underestimates the relative contact area \citep{hyun2004finite,Dapp14,Yastrebov15,xu2024new}, particularly in low-to-medium load ranges (see Fig. \ref{fig:Fig_6}(a)). \citet{Yang08} proposed the following scheme to improve the accuracy of the pressure variance:
\begin{equation}\label{eq:Correction_scheme_1}
\frac{d\text{Var}(p)}{d\zeta} = S(\bar{p}, \zeta) \frac{dV}{d\zeta}, 
\end{equation}
and the fudge factor $S(\bar{p}, \zeta)$ is a quadratic function of $A^*$, $S(\bar{p}, \zeta) = \gamma + (1 - \gamma) \left[A^*(\bar{p}, \zeta)\right]^2$, where $\gamma \in [0.42, 0.48]$ \citep{Almqvist11, Dapp14, Wang17, Afferrante18}. \citet{Wang17} proposed a similar fudge factor using a quartic polynomial of $A^*$, $S(\bar{p}, \zeta) = \gamma - \frac{2}{9} \left[A^*(\bar{p}, \zeta)\right]^2 + (\frac{11}{9} - \gamma) \left[A^*(\bar{p}, \zeta)\right]^4$, where $\gamma = 5/9$. Since solving $\text{Var}(p)$ at $\zeta$ requires the entire history of $A^*(\bar{p}, 1 \leq \zeta' < \zeta)$ to be solved iteratively in advance, it significantly increases the computational complexity of Persson's theory, particularly when determining $P_0(p, \zeta)$ and $A^*(\bar{p}, \zeta)$. \citet{xu2024new} provided an alternative scheme that directly corrects $\text{Var}(p)$:
\begin{equation}
\text{Var}(p) = S(\bar{p}, \zeta) V(\zeta), 
\end{equation}
where $S(\bar{p}, \zeta) = \gamma + (1 - \gamma) A^*(\bar{p}, \zeta)$ and $\gamma = 0.45$. Since $S(\bar{p}, \zeta)$ only depends on the ``present" magnification $\zeta$, the computational complexity of $A^*$ and $P_0(p, \zeta)$ is significantly reduced. By simply replacing $V(\zeta)$ with $\text{Var}(p)$ in the expressions for $P_0(p, \zeta)$ and $A^*$ (for instance, Eqs. \eqref{eq:pPDF_nominallyflat}, \eqref{eq:rel_area}, \eqref{eq:PDF_p_general}), the accuracy of Persson's theory is notably improved, particularly for the relative contact area $A^*$ (see Fig. \ref{fig:Fig_6}). 

On the other hand, the elastic energy of the manifold depends on the wavelength $\lambda$ of the surface undulation. In most cases, this dependence follows a power law of $\lambda^{-n}$, where $n$ ranges from $n=0$ (an Einstein solid or Winker-Fuss foundation) to $n=4$
(freely suspended sheets), with $n=1$ corresponding to the widely
used elastic half-space. By investigating the adhesive-less contact between these different elastic bodies and randomly rough surfaces, \citet{muser2021elastic} elucidated the potential origin of the fudge factor.

Recently, \citet{zhou2025} applied the physics-informed neural network (PINN) framework \citep{raissi2019physics} in Persson's theory of contact. Trained on numerical results at lower $\zeta$ ranges, the PINN accurately predicts stress distributions and contact areas at larger $\zeta$ ranges while maintaining physical consistency through Persson's diffusion equation. The corrections to the diffusion coefficient, along with the boundary conditions and initial conditions, can be effectively obtained using the PINN for both full and partial contact scenarios, demonstrating strong predictive capabilities for large-scale rough surface contacts.

\subsection{Evolution of PDF of interfacial gap with magnification}

Persson's theory commonly assumes that the purely normal contact problem is load-driven, i.e., the mean normal pressure, $\bar{p}$, is constant and independent of $\zeta$. Consequently, the drift term in the diffusion equation of $P_0(p, \zeta)$ vanishes \citep{xu2024persson}, while the mean interfacial gap $\langle g \rangle(\zeta)$ is magnification-dependent. The determination of the evolution of the PDF of the interfacial gap requires $\langle g \rangle(\zeta)$ in advance. An initial attempt was proposed by \citet{Persson07} using a thermodynamic approach, i.e., $\bar{p} = \frac{1}{A_{\text{n}}}\partial U_{\text{el}}/\partial \langle g \rangle$, where $U_{\text{el}}$ is the elastic strain energy of the deformed interface over the nominal contact area $A_{\text{n}}$. It is straightforward to solve the limiting form of $U_{\text{el}}$ at complete contact in the frequency domain \citep{Persson07}. At partial contact, Persson \citep{Persson07} provided a closed-form solution of $U_{\text{el}}$ that is weighted by the relative contact area at each magnification as follows: 
\begin{equation}
U_{\text{el}}(\bar{p}, \zeta) = \frac{\pi}{2} A_{\text{n}} E^* \int_{q_{\text{l}}}^{\zeta q_{\text{l}}} q^2 C(q) A^*(\bar{p}, q/q_{\text{l}}) S(\bar{p}, q/q_{\text{l}}) dq.
\end{equation} 
A rigorous proof of the above expression was given by \citet{Persson08}.

\begin{figure}[h!]
  \centering
  \includegraphics[width=14cm]{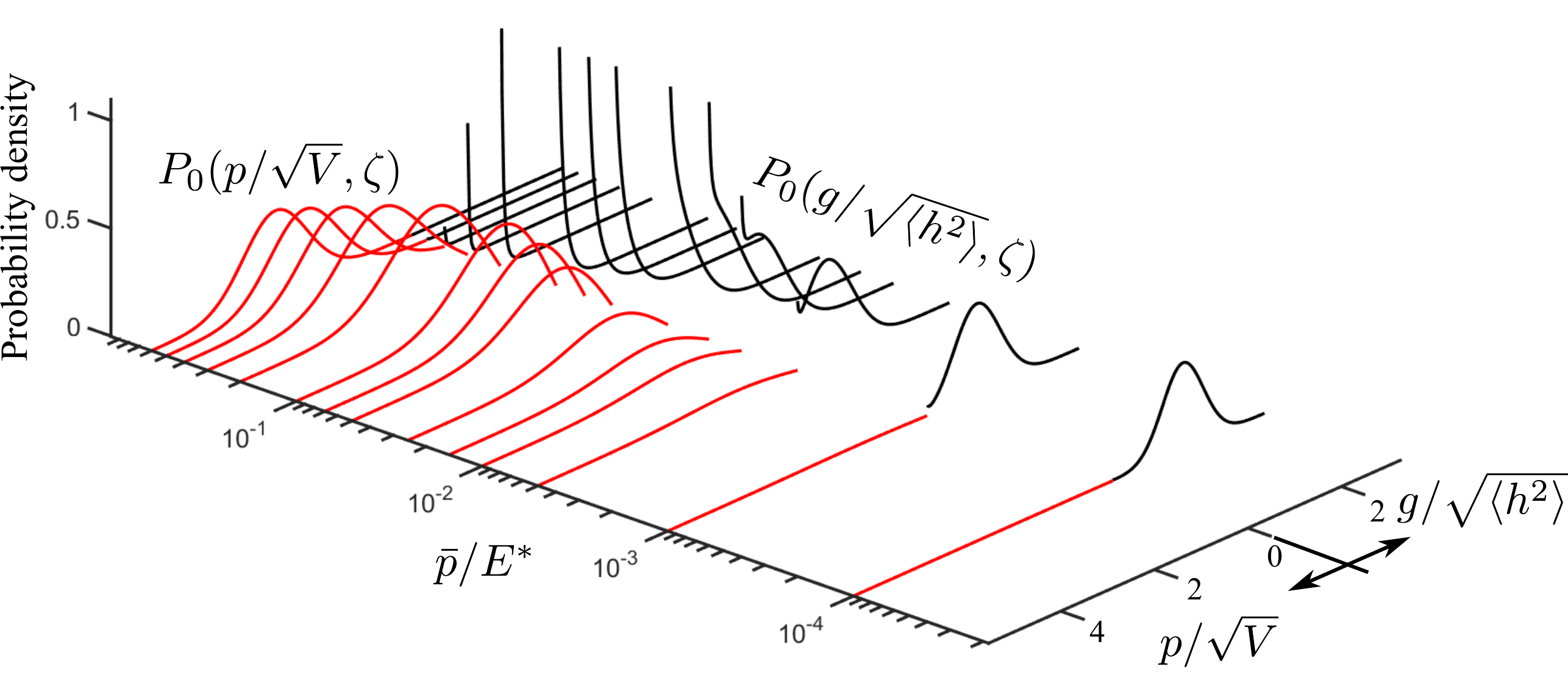}
  \caption{Evolution of $P_0(g, \zeta)$ and $P_0(p, \zeta)$ with $\zeta$. Reproduced with permission from \citet{xu2024stochastic}. Copyright 2024 Elsevier.}\label{fig:Fig_7}
\end{figure}

The PDF of the interfacial gap $P(g, \zeta)$ takes a composite form complementary to that of $P(p, \zeta)$, i.e., $P(g, \zeta) = A^* \delta(g) + (1 - A^*) P_0(g, \zeta)$, where $P_0(g, \zeta)$ denotes the PDF of the interfacial gap within the non-contact area. \citet{Yang08} did not derive the partial differential equation for $P_0(g, \zeta)$ but directly formulated $P_0(g, \zeta)$ based on its fundamental definition (i.e., the derivative of the cumulative probability with respect to $g$). Two variants of Persson's theory, following this approach, have been developed by \citet{Almqvist11} and \citet{Afferrante18}, respectively. \citet{JOE2023105397} developed a partial differential equation to characterize the evolution of $P_0(g, \zeta)$ with $\zeta$, where the non-adhesive interaction follows a power-law type of regularized adhesive surface interaction law. Inspired by the work of \citet{JOE2023105397}, \citet{xu2024stochastic} derived the following convection-diffusion equation: 
\begin{equation}
\frac{\partial }{\partial \zeta} P_0(g, \zeta) = B_1(\zeta) \frac{\partial }{\partial g} P_0(g, \zeta) - \frac{1}{2} B_2(\zeta) \frac{\partial^2 }{\partial g^2} P_0(g, \zeta), 
\end{equation}
where the drift and diffusion coefficients are approximated by $B_1(\zeta) \approx \partial \langle g \rangle/ \partial \zeta$ and $B_2(\zeta) \approx d\langle h^2 \rangle/d\zeta$, respectively. The partial differential equation above was solved numerically using a specially developed finite difference method, ensuring that the non-negativity of $P_0(g, \zeta)$ is strictly satisfied (see Fig. \ref{fig:Fig_7}). The same convection-diffusion equation was re-derived by \citet{zhou2025a} and  \citet{Xu2025Revisiting} following a method originally proposed by \citet{Persson01}. By training the lower range magnification results in the PINN, it is possible to accurately predict the PDF of gap distribution at high range magnification. The diffusion and drift coefficients, along with the boundary and initial conditions, can be effectively obtained. 

\subsection{Validation}
\begin{figure}[h!]
  \centering
  \includegraphics[width=16cm]{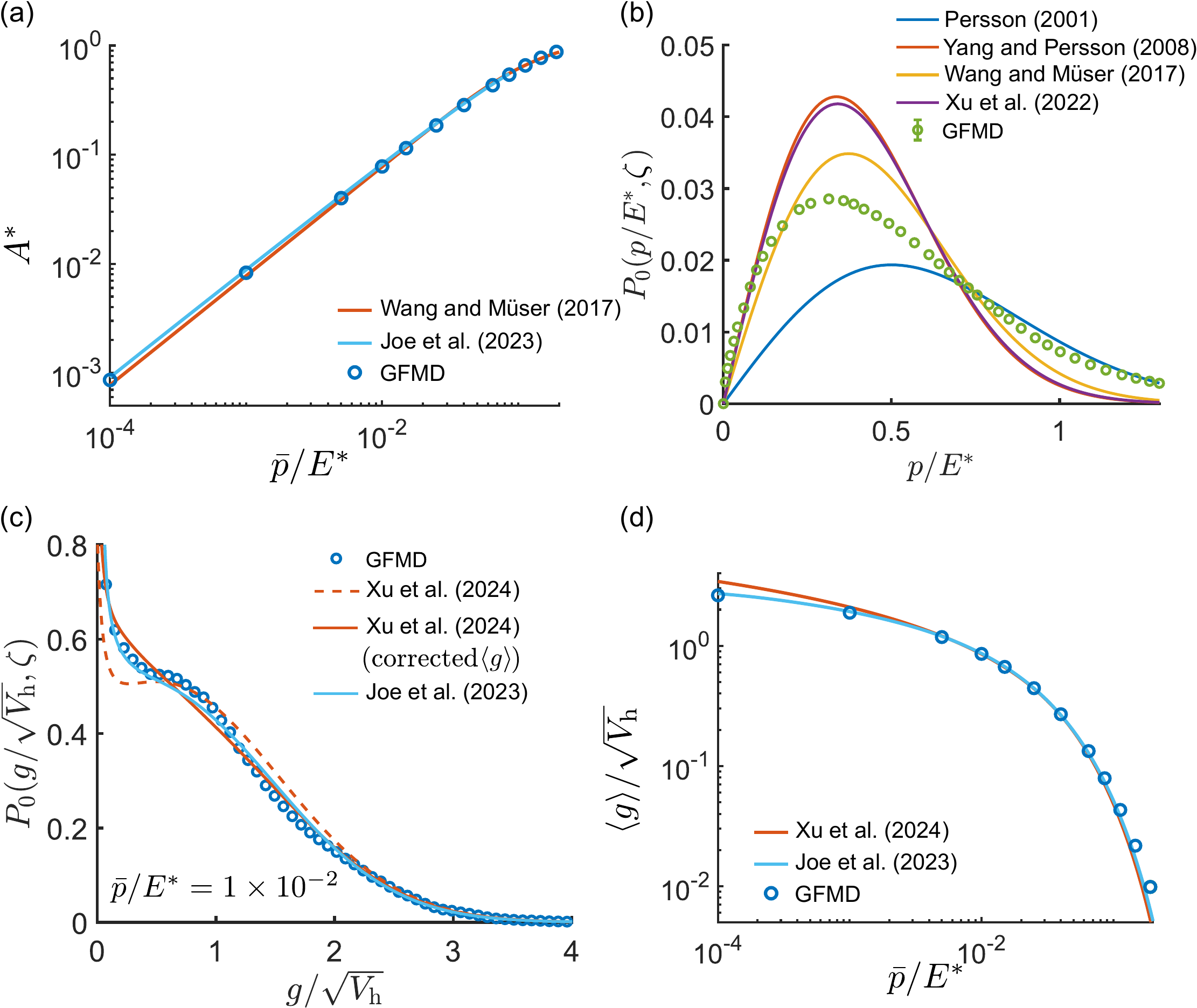}
  \caption{Comparison of Persson's theory with numerical models: (a) relative contact area; (b) PDF of the contact pressure; (c) PDF of the interfacial gap ($V_{\text{h}}$ is the variance of rough surface height); (d) average interfacial gap; $\sqrt{V_{\text{h}}}$ is the root mean square rough surface height. 
(b) is reproduced with permission from \citet{xu2024new}, (a, c, d) are reproduced with permission from \citet{xu2024stochastic}. Copyright 2017 Springer Nature and 2024 Elesvier.}. \label{fig:Fig_8}
\end{figure}

The major predictions of Persson's theory of purely normal, non-adhesive, elastic contact, namely the relative contact area \citep{putignano2012influence, prodanov2014contact,yastrebov2012contact,Yastrebov15,xu2024new}, the PDF of the contact pressure \citep{hyun2004finite, Yang06, Yang08, hyun2007elastic, campana2008elastic, putignano2012influence, Dapp14, Wang17}, the PDF of the interfacial gap \citep{Almqvist11,Afferrante18,xu2024stochastic}, the average interfacial gap \citep{Yang08,xu2024stochastic,prodanov2014contact}, the contact stiffness \citep{pastewka2013finite}, and the elastic strain energy of the deformed interface \citep{Dapp14,xu2024new}, have been extensively validated against various numerical models, including finite element \citep{hyun2004finite,hyun2007elastic}, boundary element \citep{putignano2012influence,yastrebov2012contact,Yastrebov15}, molecular dynamics \citep{Yang06}, and Green's function molecular dynamics models \citep{campana2008elastic,Dapp14,prodanov2014contact, Wang17, xu2024stochastic}. With appropriate corrections, nearly all important mechanical interfacial properties mentioned above, predicted by several variants of Persson's theory, have shown quantitative agreement with numerical results. Fig. \ref{fig:Fig_8} shows a summary of some representative comparisons.

As pointed out by \citet{prodanov2014contact}, all numerical contact mechanics models are developed over a finite computational domain of size $L$, with a finite magnification of $\zeta = \lambda_{\text{l}}/\lambda_{\text{s}}$ and a finite element size (or lattice resolution in GFMD) $a$. Persson's theory of contact, which models the interaction of bandwidth-limited rough surfaces, should be applied within thermodynamic and continuum limits (TC limits), i.e., $1/L \to 0$ and $a/\lambda_{\text{s}} \to 0$. Therefore, without careful post-processing of the numerical results, comparisons may be unjustified. For instance, in early studies, $P_0(p, \zeta)$ predicted by finite element \citep{hyun2004finite, hyun2007elastic} and molecular dynamics models \citep{Yang06} often deviated from the theoretical predictions of Persson's theory, especially in low-pressure regimes where the absorbing boundary condition was not accurately captured. This discrepancy arises from the fact that $\lambda_{\text{s}}/a$ is not sufficiently large in those numerical models. \citet{prodanov2014contact} proposed an extrapolation method to predict the interfacial mechanical properties at TC limits. To obtain the relative contact area at the continuum limit, \citet{yastrebov2017accurate} proposed a correction scheme for post-processing the contact patch distribution obtained from the boundary element method using constant elements (where each element assumes a uniform pressure distribution) so that the corrected relative contact area is independent of element size.

\begin{figure}[h!]
  \centering
  \includegraphics[width=16cm]{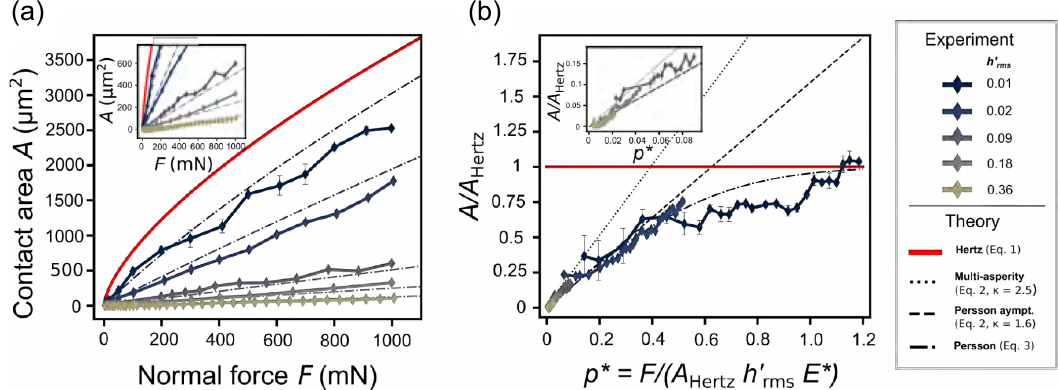}
  \caption{Experimental validation of Persson's theory. Reproduced with permission from \citet{terwisscha2024elastic} under the terms of the CC-BY 4.0 License (https://creativecommons.org/licenses/by/4.0/). Copyright 2024 Elsevier.}\label{fig:Fig_9}
\end{figure}

Persson's theory has been quantitatively validated in several independent experiments. In the well-known Contact Mechanics Challenge \citep{muser2017meeting}, \citet{bennett2017contact} optically measured the relative contact area between a nominally flat rough PDMS (Polydimethylsiloxane) surface and a glass plate  under various normal loads. Multiple PDMS rough surfaces were fabricated by molding against a 3D-printed PMMA (Polymethyl methacrylate) plate with an isotropic surface texture that was numerically generated with a known PSD. This method of preparing rough surface samples guarantees high repeatability of the surface topography, as well as the PSD \citep{perris20233d}, which is essential for a fair comparison between the tests and Persson's theory. The relative contact area predicted by Persson's theory quantitatively matches the optical measurements, albeit with a slight underestimation. This discrepancy may be attributed to the fact that the chosen variant of Persson's theory does not account for surface adhesion. \citet{terwisscha2024elastic} optically measured the contact area of a brittle material contact pair between a soda-lime glass sphere and a glass coverslip, where the effect of adhesion can be largely ignored due to the stiff substrates of both surfaces (see Fig. \ref{fig:Fig_9}). The spherical surfaces were roughened using various grits of sandpaper to achieve different levels of root mean square roughness ($S_{\text{q}}$). As $S_{\text{q}}$ increases, $A^*(\bar{p})$ transitions from a nominally flat rough contact solution to a classic Hertzian solution. Equation \eqref{eq:rel_area} shows excellent agreement with the measured data for all selected values of $S_{\text{q}}$, where the nominal contact area of a circular shape is determined by Hertzian theory. 

\section{Non-adhesive inelastic contact}
Persson's theory of contact can be applied not only to elastic contact but also to inelastic contact, including elastoplastic and viscoelastic interactions. 

\subsection{Elastoplastic contact} 
Regarding elastoplastic contact, \citet{Persson01} and \citet{Persson01b} made the initial attempt to incorporate this nonlinear phenomenon by assuming a constant pressure distribution within the contact area ($p = H_0$, where $H_0$ is a constant hardness). This approach yields a composite form of $P(p, \zeta)$, as follows: \citep{Persson01b,Xu22}
\begin{equation}
P(p, \zeta) = P_0(p, \zeta) + A_{\text{pl}}^* \delta(p - H_0) + (1 - A^*) \delta(p),
\end{equation}
where $P_0(p, \zeta)$ is the magnification-dependent PDF of the contact pressure within the elastically deformed contact area; $A^* = A_{\text{el}}^* + A_{\text{pl}}^*$, where $A_{\text{el}}^*$ and $A_{\text{pl}}^*$ are the relative elastic and plastic contact areas, respectively. The evolution of $P_0(p, \zeta)$ with $\zeta$ follows the same diffusion equation as given in Eq. \eqref{eq:Diffusion} with absorbing boundary conditions $P_0(p = 0, \zeta) = P_0(p = H_0, \zeta) = 0$, which has been rigorously proven by \citet{Xu22}. 
The closed-form expressions for $P_0(p, \zeta)$, $A_{\text{el}}^*$, and $A^*_{\text{pl}}$ were initially derived by Persson as infinite sums of sinusoidal terms \citep{Persson01b}. 
When $\bar{p} <H_0$, the relative elastic contact area monotonically decreases as $\zeta$ increases. 
The ``leakage" of $P_0(p, \zeta)$ occurs at $p = 0$ and $p = H_0$ increases the relative non-contact area ($1 - A^*$) and the relative plastic contact area ($A_{\text{pl}}^*$), such that $A^*$ converges to a constant at larger $\zeta$, where the deformation is predominantly plastic \citep{lambert2025competition}. 
This work has been criticized by \citet{borodich2002comment} argues that Persson's model is based purely on geometry and lacks equations of elasticity and plasticity. 
In fact, the elasticity equation is indeed included in Persson's theory in terms of the diffusion coefficient $B_2(\zeta)$ (Eq. \eqref{eq:Diffusion_approximate}), and the plasticity equation is simply approximated by an elastic-perfectly plastic constitutive law. 

\begin{figure}[h!]
  \centering
  \includegraphics[width=16cm]{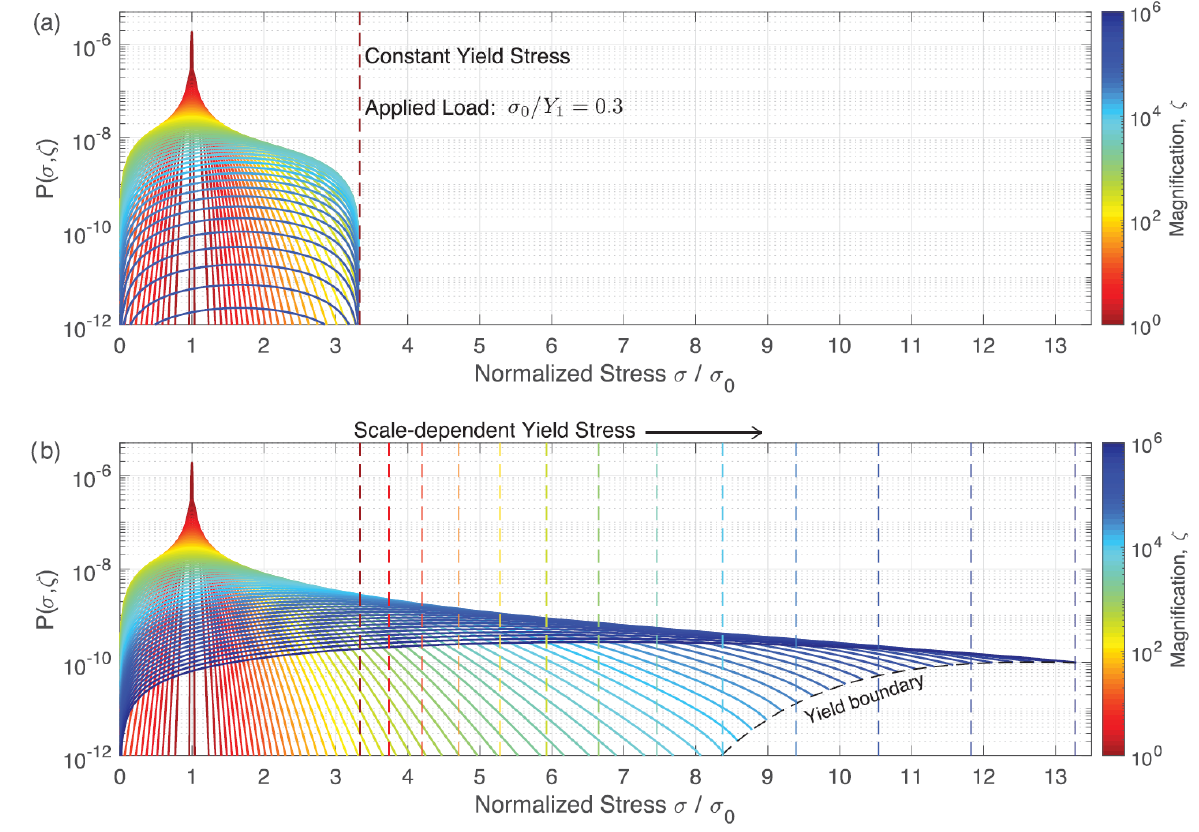}
  \caption{Evolution of PDF of contact pressure with magnification: (a) constant hardness and (b) scale-dependent hardness. It should be noted that $P(\sigma, \zeta)$, $\sigma_0$, $Y_1$ in both (a) and (b) denote, respectively, $P_0(p, \zeta)$, $\bar{p}$ and $H(\zeta = 1)$. For consistency with the original text, we did not adjust the axis label accordingly. Reproduced with permission from \citet{lambert2025competition}, Copyright 2025 American Physics Society.}\label{fig:Fig_10}
\end{figure}

\citet{Xu22} simplified the expression of $P_0(p, \zeta)$ into a sum of three Gaussian distributions, represented as follows:
\begin{equation}\label{eq:Persson_elastoplastic}
P_0(p, \zeta) =  \frac{1}{\sqrt{2 \pi V(\zeta)}} \left\{ \exp\left[ -\frac{(p - \bar{p})^2}{2 V(\zeta)} \right] - a \cdot \exp\left[ -\frac{(p + \bar{p})^2}{2 V(\zeta)} \right] - b \cdot \exp\left[ -\frac{(p - 2 H_0 + \bar{p})^2}{2 V(\zeta)} \right] \right\},
\end{equation}
which naturally satisfies the diffusion equation and the initial condition. The implementation of two absorbing boundary conditions at $p = 0$ and $p = H_0$ leads to closed-form expressions for $a$ and $b$ \citep{Xu22}. The expressions for $a$ and $b$ provided by \citet{Xu22} are not in their simplest forms; the following simplified versions are:
\begin{align}
a &= \left\{ 1 - \exp[-2(H_0 - \bar{p}) H_0/V(\zeta) ]\right\}/\left[ 1 - \exp(-2H_0^2/V(\zeta)\right], \label{eq:a_Persson}\\
b &= 1 - a \cdot \text{exp}\left(-2 \bar{p} H_0/V(\zeta) \right). 
\end{align}
For a rough profile contact, \citet{venugopalan2019plastic} developed a fudge factor to improve the accuracy of the predicted $A^*$ based on the numerical results obtained from the Green's function dislocation dynamics (GFDD) simulation. 
\citet{ciavarella2024new} extended Persson's theory to incorporate a scale-dependent hardness model. A new plasticity index has been derived that qualitatively explains the asperity persistence observed in indentation tests. 
\citet{lambert2025competition} further extended Persson's theory by including the scale-dependent hardness $H(\zeta)$. Based on load equilibrium and probability conservation, they derived a partial differential equation in terms of $P_0(p = H(\zeta), \zeta)$, while $P_0(p = 0, \zeta) = 0$. The corresponding $P_0(p, \zeta)$ has been solved numerically using a finite difference method (see Fig. \ref{fig:Fig_10}). Importantly, they analytically derived a parameter to quantify the competing effects of the multi-scale nature of surface roughness and the hardness scaling law on plastic deformation.

For light loads and relatively smooth contact, Persson's theory shows good agreement with the results of the GFDD model \citep{venugopalan2019plastic}. Due to a lack of strain hardening and size-dependent plasticity, $P(p, \zeta)$ predicted by Persson's theory occupies a relatively narrow pressure range and presents an unrealistic pulse-like peak at $p = H_0$. In contrast, $P(p, \zeta)$ predicted by the GFDD model is continuous and spans a broader pressure range, even extending beyond $p = H_0$.
\citet{patil2025roughness} recently conducted in-situ measurements of the real contact area and average interfacial gap of the Polymethyl methacrylate (PMMA)-Germanium contact interfaces. They demonstrated that Persson's theory of elastoplastic contact can accurately predict the scaling law of the area-to-load relationship within both the elastic and fully plastic ranges.

\subsection{Viscoelastic contact} 
In his seminal work on tire-road interaction \citep{Persson01}, Persson derived a general form of the surface displacement components of a viscoelastic half-space in response to normal and shear tractions. The first attempt to solve the purely normal viscoelastic contact problem was made by \citet{persson2004contact} who based his solution on elastic results in the frequency domain and subsequently transformed it back to the time domain. Persson provides a closed-form solution for the magnification- and time-dependent relative contact area during creep. This method is known in solid mechanics as the elastic-viscoelastic correspondence principle \citep{christensen2013theory}, which can only be applied when the contact area increases monotonically \citep{greenwood2010contact}. The conventional theory of contact proposed by \citet{persson2004contact} is applicable only to specific cases where a viscoelastic half-space is in contact with a rigid rough surface. \citet{Scaraggi15} removed this restriction and extended Persson's theory to a rough-on-rough scenario. The relative contact area has been formulated in terms of the power spectral density of two rough surfaces. \citet{Papangelo21} extended Persson's theory of viscoelastic contact to more complex loading conditions, including step pressure loading, step displacement loading, and oscillatory loading. With oscillatory loading, they showed that the relative contact area reaches its maximum not at the peak load but during the unloading stage, where the conventional correspondence principle \citep{persson2004contact} can no longer be considered valid. They analytically solved the time evolution of the relative contact area and average interfacial gap during the first oscillatory cycle using Ting's and Greenwood's modified correspondence principle \citep{ting1966contact,greenwood2010contact}. 

\section{Adhesive contact}
Before the creation of Persson's theory of contact, adhesive contact problems were predominantly solved analytically using multi-asperity contact models \citep{fuller1975effect,ciavarella2017effect} based on the Johnson-Kendall-Roberts (JKR), Derjaguin-Muller-Toporov (DMT), and Maugis-Dugdale (MD) adhesion models at the asperity level \citep{johnson1971surface}. 
For adhesive contact in the JKR limit, where surface adhesion is neglected outside the contact area, \citet{persson2002adhesion} utilized the same diffusion equation as presented in Eq. \eqref{eq:Diffusion} to characterize the evolution of the PDF of the adhesive traction with respect to magnification. 
The initial and boundary conditions are nearly identical to those of non-adhesive contact, with the exception of the absorbing boundary condition at a critical tensile traction, which is as follows:
\begin{equation}
P_0(p = -\sigma_{\text{a}}(\zeta), \zeta) = 0.
\end{equation}
If $\sigma_{\text{a}}$ is independent of $\zeta$, the closed-form expression of $P_0(p, \zeta)$ is readily available by slightly modifying the mirrored Gaussian distribution in Eq. \eqref{eq:pPDF_nominallyflat}. If $\sigma_{\text{a}}(\zeta)$ is magnification-dependent, the expression of $P_0(p, \zeta)$ in integral form can be derived after applying the change of variable \citep{persson2002adhesion}. The critical tensile traction, $\sigma_{\text{a}}(\zeta)$, is defined as the maximum traction acting on the penny-shaped unbounded regions with a magnification-dependent size, just before it jumps into contact with the mating surface \citep{persson2002adhesion}. \citet{carbone2009adhesive} extended Persson's adhesion model to the contact of rough profiles under plane strain conditions. \citet{Wang17} proposed corrections to the aforementioned Persson's adhesive contact model \citep{persson2002adhesion}, where the diffusion coefficient, $B_1(\zeta)$, was modified to be locally pressure-dependent using a similar fudge factor for correcting the relative contact area \citep{Wang17}. 

For adhesive contact in the DMT limit, the total surface traction is the superposition of the elastic non-adhesive contact pressure within the contact area, $p(x, y)$, and the tensile traction outside the contact area, $p_{\text{a}}(x, y)$. The PDF of the latter traction can be estimated based on the intermolecular/atomic attraction law and the interfacial gap between the outermost molecular/atomic layers of the two surfaces \citep{persson2014theory, xu2024stochastic}. Additionally, a scale-dependent Tabor's parameter has been developed by \citet{persson2014theory} to characterize the transition of rough surface adhesion between the JKR and DMT limits. For a general adhesion problem that lies between the JKR and DMT limits, \citet{Joe17} and \citet{Joe18} formulated the evolution of the PDF of the interfacial gap with magnification using the Chapman-Kolmogorov equation, which they later extended to a differential form \citep{JOE2023105397}. In the model of Joe et al., the interface is in separation everywhere, with no distinction made between the contact and non-contact areas. The adhesive traction and interfacial gap are coupled using Lennard-Jones potential. 

\citet{Persson01a} defined an effective interfacial energy $\gamma_{\text{eff}}$ that quantitatively characterizes the impact of a rough surface on adhesion. This effective interfacial energy consists of two components: the surface energy and the elastic strain energy:
\begin{equation}\label{eq:gamma_eff}
\gamma_{\text{eff}} A^*(\zeta) = \Delta \gamma A^*(\zeta) \eta - U_{\text{el}}(\bar{p}, \zeta)/A_n,
\end{equation}
where $\Delta \gamma$ denotes the work of adhesion; $\eta > 1$ is a scaling factor indicating that the bonded area is larger than the nominal contact area, $A^*$. The effective interfacial energy exhibits a generally non-monotonic trend, initially increasing before subsequently decreasing as magnification increases \citep{Persson01,persson2002adhesion}. This observation suggests that both adhesion strengthening and weakening can be modulated by the geometrical and multi-scale features of the rough topography. Equation \eqref{eq:gamma_eff} quantitatively explains the adhesion paradox, which states that adhesion is negligible at the macro-scale \citep{persson2004nature}. 

\begin{figure}[h!]
  \centering
  \includegraphics[width=16cm]{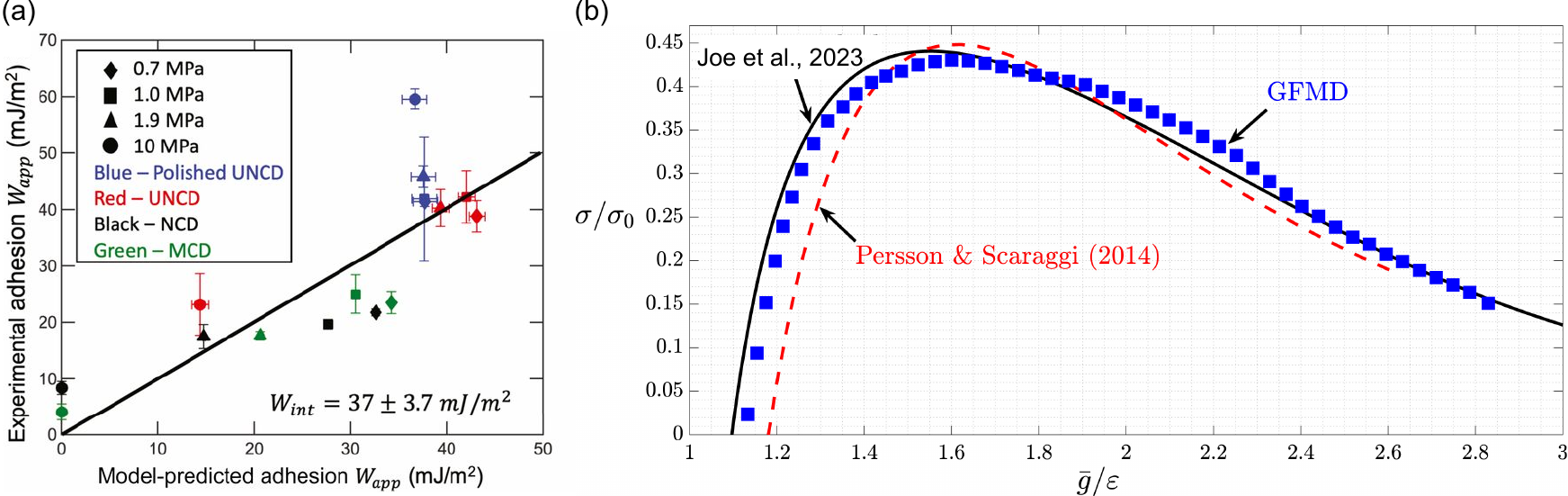}
  \caption{Validation of Persson's theory of adhesive contact: (a) effective work of adhesion against experimental results and (b) force-approach curve against numerical results. (a) is reproduced with permission from \citet{dalvi2019linking}. (b) is reproduced with permission from \citet{JOE2023105397}.}\label{fig:Fig_11}
\end{figure}

\citet{carbone2009adhesive} developed a plane strain boundary element method for simulating adhesion between rough profiles. Persson's theory qualitatively predicts a linear relationship between the relative contact area and the normal load, similar to the numerical model; however, it underestimates the relative contact area by almost 50$\%$. Consequently, the PDF of normal traction, which includes both compressive and tensile pressures, predicted by Persson's theory is also underestimated. \citet{carbone2009adhesive} showed that a mirrored (double) Gaussian distribution of normal traction can accurately fit the numerical results by tuning the $\sigma_a$ and the diffusion coefficient. \citet{dalvi2019linking} conducted in-situ measurements probing the adhesive behavior of rough PDMS spheres on a polycrystalline diamond substrate. The measured effective work of adhesion shows good agreement with that predicted by Persson's theory \citep{Persson01a}, see Fig. \ref{fig:Fig_11}(a). Variants of Persson's theory developed  \citep{Joe17,Joe18,JOE2023105397} accurately predict $P(g, \zeta)$, $P(p, \zeta)$, $\langle g \rangle$ to $\bar{p}$ relation with good agreement with those obtained from two independent adhesive numerical models, see Fig. \ref{fig:Fig_11}(b).  

\section{Other fields of applications}
\subsection{Friction}

Persson's theory of contact has been extensively applied to study rubber friction, which arises from two sources: the viscoelastic deformation of the rubber and adhesive shearing at the contact area \citep{fortunato2017dependency}. This theory was originally developed to model a viscoelastic half-space in sliding contact with a rigid, isotropic, rough surface \citep{Persson01}, making it potentially applicable in tire-road interactions. Persson initially assumed that the frictional energy induced by shearing is equal to the strain energy stored in the deformed interface, without being dissipated through frictional heating. Consequently, the coefficient of sliding friction at steady state has been formulated using an energetic approach. A simplified form of the coefficient of sliding friction has been obtained by \citet{ciavarella2018simplified}. General formulations of the coefficient of friction have been extended by \citet{scaraggi2015general} and \citet{scaraggi2016effect} to cover more complex rough-on-rough contact scenarios and layered contact. 
\citet{mokhtari2016transversely} extended Persson's friction model from linear viscoelastic materials to transversely isotropic viscoelastic materials. 
The temperature-dependent mechanical properties of materials have been incorporated into Persson's theory \citep{persson2006rubber}, which quantitatively predicts that frictional heating can lead to a decrease of the coefficient of sliding friction, particularly when the sliding velocity exceeds $0.01$ m/s. 
\citet{carbone2009contact} extended Persson's original work to model the friction between anisotropic rough surfaces. The relative contact area shows a weak dependency on the sliding direction, while the coefficient of friction is strongly dependent on the sliding direction. 
\citet{mokhtari2014friction} introduced the adhesive term into Persson's theory of contact, which quantitatively characterizes the shear stress induced by adhesion and the shearing of a fluid-like thin film molecular layer. 
Persson's theory has been extended by \citet{persson2002theory}  to non-stationary sliding contact to reproduce the rate- and state-dependent friction law observed in friction tests. 
%


\subsection{Electrified interface}
\citet{barber2003bounds} proved that the electrical contact resistance of an elastic rough contact interface has a linear relationship with normal contact stiffness. Building upon this conclusion, \citet{persson2022electric} derived a closed-form solution for electrical contact resistance when the normal load is infinitesimally small. Persson's theory has also been extended to examine electroadhesive contact between two conducting surfaces \citep{Persson18, persson2019electroadhesion, persson2021general}. By formulating the electrostatic attraction based on the PDF of the interfacial gap, \citet{Persson18} derived a nonlinear equation for normal load. His electroadhesive theory quantitatively predicts the evolution of the adhesive force, the average interfacial gap, and the relative contact area with applied voltage in the steady state \citep{Persson18}. The response of the adhesive force to the transient applied voltage of various shapes has also been studied \citep{persson2019electroadhesion}. \citet{persson2020role} applied his theory of contact to solve the contact electrification problem, assuming that the driving force of the charged species is the electric field induced by the flexoelectric effect \citep{mizzi2019does}. 

\subsection{Seal}
Because of the inevitable uneven topography at the seal interface, there is no perfect static seal that achieves zero leakage. \citet{persson2008theory} developed a theoretical approach to estimate the leakage rate of a static seal. A critical magnification $\zeta_{\text{c}}$ exists, beyond which the contributions of roughness components with smaller wavelengths to the leakage rate become negligible. Based on critical junction theory, $A^*(\zeta = \zeta_{\text{c}}) \approx 0.4$. Using Persson's theory of contact, one can solve for the critical magnification ($\zeta_{\text{c}}$) and the corresponding average interfacial gap ($\langle g \rangle(\zeta_{\text{c}})$). The leakage rate of the static seal subjected to a pressure difference of $\Delta p$ can be formulated as 
\begin{equation}
\dot{Q}= \frac{\langle g \rangle^3(\zeta_{\text{c}})}{12 \eta} \Delta p,  
\end{equation}
where $\eta$ represents the fluid viscosity. A similar formulation has also been developed by \citet{bottiglione09}. 
For metallic seals, \citet{persson2016leakage} showed that plastic deformation can reduce the leakage rate, particularly when the surface is rougher with a larger root mean square roughness. \citet{deng2023novel} experimentally demonstrated the validity of the leakage rate formulation of Persson's theory while considering plastic deformation. 

The critical junction theory assumes that the leakage rate is determined by the flow resistance through the critical junction.
On the other hand, when rough surfaces are in contact, numerous flow channels form at the interface.
Based on this observation, Persson applied the 2D Bruggeman effective medium theory to calculate the leakage rate resulting from the network of these flow channels.
A constant effective flow conductivity is utilized to replace the locally and spatially varying conductivity, thus the leakage rate can be expressed as
~\citep{Persson2010JPCM}
\begin{equation}
  \dot Q = \sigma_{\mathrm{eff}} \Delta p,
\end{equation}
where $\sigma_{\mathrm{eff}}$ denotes the effective flow conductivity and can be determined by
\begin{equation}
  \int_0^{\infty} \frac{2\sigma_{\mathrm{eff}}}
  {\sigma_{\mathrm{eff}} + g^3/12 \eta} P_0(g, \zeta) dg = 1,
\end{equation}
where $g$ is the local interfacial gap, and $P_0(g, \zeta)$ denotes the PDF of the interfacial gap distribution, which can be determined by Persson's theory \citep{Almqvist11,Joe22,xu2024stochastic}.

For isotropic rough surfaces, when the interfacial separation becomes very small, the effective medium theory and the critical junction theory yield almost identical results.
However, the critical junction theory is significantly simpler to implement and requires considerably less computational time, even under high squeezing pressures. In contrast, under such conditions, the effective medium theory tends to become unstable.
Furthermore, by considering the scale-dependent Peklenik number, both the effective medium theory and the critical junction theory can be further extended to investigate leakage problems of anisotropic rough surfaces~\citep{Persson2012EPJE,Persson2020EPJE,Wang2020TL}.

\subsection{Wear}
\citet{persson2025rubber} have recently developed a wear model for rubber tire compounds in sliding contact with a rough concrete surface. Based on the assumptions that (1) rubber asperities of a certain size is removed under a critical shear stress and (2) the ratio of shear stress to normal stress remains independent of the sliding coefficient of friction, the size of the worn surface associated with the wear particle, at a specific magnification, can be analytically estimated. By summing the wear particles from all length scales, an Archard-like model \citep{archard1953contact} has been developed to predict the total wear volume per unit sliding distance over a nominal contact area. The model shows good quantitative agreement with published experimental data \citep{xu2025rubber}. \citet{Ciavarella25some} further simplified the wear model of Xu and Persson by demonstrating that it is essentially the same as the Archard model, which has a closed-form solution for the wear coefficient that requires prior experimental calibration. 

\section{Future work}

\subsection{More fundamental studies}
In recent years, the author has observed a growing interest in applying Persson's theory of contact to diverse applications, including those outside conventional tribological fields (e.g., earthquakes \citep{lambert2025competition} and rock engineering \citep{lang2015hydraulic,kling2018numerical}. However, fundamental studies of Persson's theory of contact receive little attention and is primarily conducted by Persson and a few of his close collaborators. 
This limited focus is largely due to the fact that the original derivation of Persson's theory (see Eqs. (B1)-(B5) in Appendix B of \citep{Persson01}) is oversimplified for tribologists and solid mechanicians who lack a background of statistical mechanics. 
Few attempts \citep{Manners06,xu2024persson,Xu2025Revisiting} have been made to ``decipher" this theory. Among these, \citet{Manners06} were the first to provide important comments on the physical meaning of the diffusion coefficient, as well as the derivation of the diffusion equation. A revisit of Eqs. (B1-B5) has been conducted by \citet{Xu2025Revisiting}, which serves as complementary material for understanding Persson’s original work. \citet{xu2024persson} provided a detailed tutorial illustrating how to derive the diffusion equation using stochastic process theory, which has been proven to be an effective tool \citep{Joe17,Joe18,Joe22,JOE2023105397,xu2024stochastic} for probing the multi-scale features of the random mechanical properties over the rough contact interface. More fundamental studies on Persson's theory of contact from different perspectives are always welcome to assist researchers who are not familiar with, or are still struggling to understand, Persson's theory. 

Nearly no studies \citep{Dapp14} have addressed the core assumption adopted in Persson's theory regarding whether the Markov process is strictly followed by the random contact pressure. Future numerical studies should thoroughly investigate the validity of this assumption in both elastic and inelastic contact problems. We should identify easily implementable approaches to abandon the no-reentry assumption. Inspired by recent work by \citet{zhou2025}, future research on Persson’s theory could rely on artificial intelligence to formulate partial differential equations with higher accuracy. 

The current variants of Persson's theory can only characterize the magnification-dependent evolution of a single random variable (e.g., contact pressure or interfacial gap) using either a partial differential equation or the Chapman-Kolmogorov equation. For certain applications, the interplay among multiple random variables is essential. For instance, frictional contact is jointly determined by the local contact pressure and shear stress \citep{ciavarella1998generalized,sahli2018evolution, zhang2024non}; the normal traction and interfacial gap are coupled through the Lennard-Jones potential in adhesion at the DMT limit \citep{persson2014theory}. Future studies of Persson's theory that incorporate multiple random variables may facilitate more advanced applications across various fields; 

The magnification used in Persson's theory remains a subject of debate across various applications. In certain applications, such as sealing \citep{persson2008theory}, the ranges can be derived analytically, as contributions from wavelengths below the threshold value is negligible. In other cases, such as friction \citep{ing2025validation}, predictions may significantly depend on the selected magnification. How to select a critical magnification for a specific application should be explicitly outlined in future studies.

Lastly, the strong similarity between the asymptotic behavior of Persson's theory and the multi-asperity contact model under both light and heavy loads suggests that these methods may share a common foundation, as indicated in Eqs. \eqref{eq:A_lowload}, \eqref{eq:Persson_solu}, and \eqref{eq:Ciavarella_solu}. It is possible that an unified contact model exists, in which Persson's theory of contact and the multi-asperity contact model serve as its asymptotes, respectively, in the ranges of heavy and light loads. This topic is of paramount importance for future studies on rough surface contact.

\subsection{More applications}
Persson's theory of contact is predominantly utilized to address the contact problem at the interface; however, several significant tribological phenomena occur within the subsurface beneath the nominal contact area, e.g., internal stress \citep{zhang2025investigation}, contact-induced bulk polarization \citep{mizzi2019does}, and plasticity \citep{venugopalan2019plastic}. \citet{muser2018internal} confirmed that the double Gaussian form in Eq. \eqref{eq:pPDF_nominallyflat} remains applicable to characterizing the PDF of internal stress at an arbitrary depth below the nominal contact area. By adjusting the diffusion coefficient, the PDF of the stress component predicted by Persson's theory aligns well with numerical results. However, its transition from the interface to the bulk cannot be adequately characterized by Persson's theory \citep{muser2018internal}. Lastly, due to the complexity of Persson's theory, an open-source computational library should be developed by experts in this field to facilitate a broader application of Persson's theory of contact in tribology and other fields. 

\section{Conclusion}
In this chapter, the historical development of multi-scale models, including Persson's theory of contact, is discussed in detail. Starting from purely normal elastic contact, the PDF of contact pressure and the interfacial gap can be derived from partial differential equations. These fundamental results have been applied across various tribological fields, including friction, sealing, wear, inelastic contact, and adhesive contact. Future topics related to Persson's theory are proposed to further advance its fundamental study and broaden its application outside tribology. We hope that this review article will assist interested readers in becoming familiar with the current state-of-the-art of Persson's theory and attract increasing attention to the development and application of Persson's theory of contact.

\begin{spacing}{1} 
    \bibliographystyle{cas-model2-names}
	\bibliography{ref}
\end{spacing}

\end{document}